\documentclass{article}
\usepackage{spconf,amsmath,graphicx}
\usepackage{subfigure}
\usepackage{amssymb}
\usepackage{xcolor}
\usepackage{booktabs}
\usepackage{diagbox}

\newcommand{\tabincell}[2]{
\begin{tabular}{@{}#1@{}}#2\end{tabular}
}
\title{Fostering the Robustness of White-Box Deep Neural Network Watermarks by Neuron Alignment}
\name{Fang-Qi Li, Shi-Lin Wang*, Senior Member, IEEE, Yun Zhu\thanks{This work was supported by the National Natural Science Foundation of China (61771310). Shi-Lin Wang is the corresponding author.}}
\address{\texttt{ \{solour\_lfq,wsl,scott0518\}@sjtu.edu.cn}\\
School of Electronic Information and Electrical Engineering, Shanghai Jiao Tong University.}
\begin{document}
%
\maketitle
\begin{abstract}
The wide application of deep learning techniques is boosting the regulation of deep learning models, especially deep neural networks (DNN), as commercial products.
A necessary prerequisite for such regulations is identifying the owner of deep neural networks, which is usually done through the watermark.
Current DNN watermarking schemes, particularly white-box ones, are uniformly fragile against a family of functionality equivalence attacks, especially the neuron permutation.
This operation can effortlessly invalidate the ownership proof and escape copyright regulations.
To enhance the robustness of white-box DNN watermarking schemes, this paper presents a procedure that aligns neurons into the same order as when the watermark is embedded, so the watermark can be correctly recognized.
This neuron alignment process significantly facilitates the functionality of established deep neural network watermarking schemes.
\end{abstract}
\section{Introduction}
\label{sec:1}
Deep learning models have made significant achievements in domains ranging from computer vision~\cite{9414465} to signal processing~\cite{9413901,9413723}.
Since deep neural networks (DNN) can provide high-quality service, they have been treating as commercial products and intellectual properties.
One necessary condition for commercializing DNNs is identifying their owners.
DNN watermark is an acknowledged technique for ownership verification (OV).
By embedding owner-dependent information into the DNN and revealing it under an OV protocol~\cite{oursijcai}, the owner of the DNN can be uniquely recognized.

If the pirated model can only be interacted as a black-box then backdoor-based DNN watermarking schemes are the only option.
They encode the owner's identity into backdoor triggers by pseudorandom mapping~\cite{zhu2020secure}, variational autoencoder~\cite{li2019prove}, or deep image watermarking~\cite{zhang2021deep}.
Adversarial samples~\cite{le2020adversarial} and penetrative triggers~\cite{oursicip} have been designed to defend against adversarial tuning and filtering.
However, in the realistic setting, an adversary can ensemble multiple DNN models or adding extra rules to invalidate backdoors.

White-box DNN watermarking schemes have better performance regarding unambiguity and forensics given unlimited access to the pirated model.
They embed the owner's identity into the model's weight~\cite{uchida2017embedding}, its intermediate outputs for specific inputs~\cite{ours}, etc.
The white-box assumption holds for many important scenarios such as model competitions, engineering testing, and lawsuits.

Despite their privileges, white-box DNN watermarking schemes are haunted by the functionality equivalence attack, in particular, the neuron permutation attack~\cite{lukas2021sok}.
The watermark is uniformly tangled with the parameters of neurons, so the adversary can invalidate it and pirate the model by permutating neurons without affecting the model's performance.

To cope with this threat and foster the robustness of white-box DNN watermarking schemes, we propose a neuron alignment framework.
By encoding the neurons and generating proper triggers, the order of neurons can be recovered.
Then the watermark can be correctly retrieved and the ownership is secured.
The contribution of this paper is threefold:
\begin{itemize}
\item We propose a DNN protection framework against the neuron permutation attack. 
To the best of our knowledge, this is the first attemp in defending such threat. 
\item By aligning neurons, the proposed framework can recover the order of neurons and can be seamlessly combined with established watermarking schemes. 
\item Experiments have justified the efficacy of our proposal.
\end{itemize}

\section{The Motivation}
\label{sec:2}
\begin{figure}[htbp]
\centering
\subfigure[The normal verification.]{
\includegraphics[width=4cm]{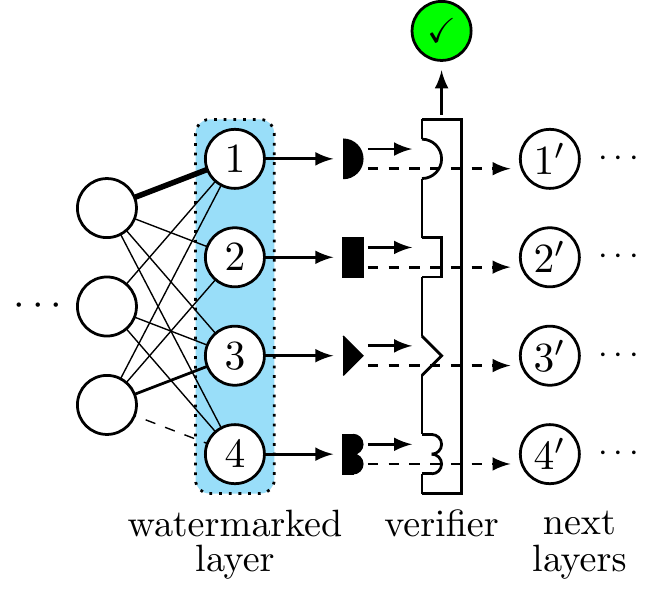}
}\subfigure[The verification under attack.]{
\includegraphics[width=4cm]{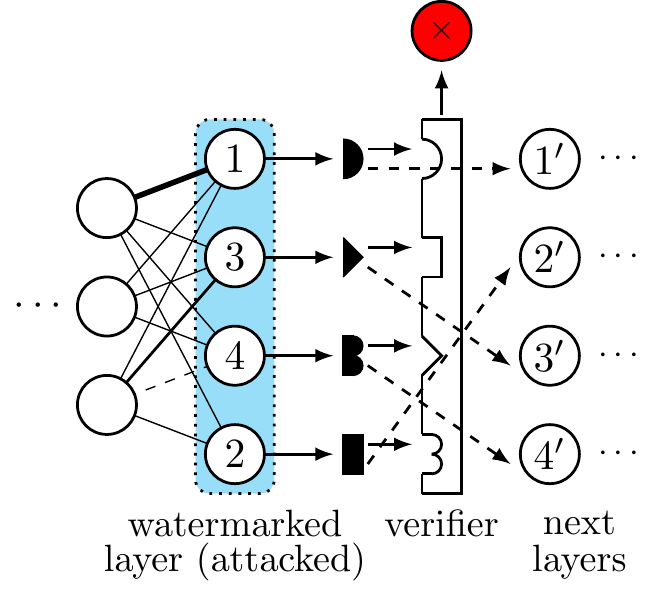}
}
\caption{A neuron permutation attack, $\textbf{P}=(2,3,4)$.}
\label{figure:threat}
\end{figure}

In OV, the verifier module takes the parameters/outputs of neurons as its input.
An adversary can shuffle homogeneous neurons (whose forms and connections to previous layers are identical) using a permutation operator $\textbf{P}$ such that the input from the verifier's perspective is no longer an identification proof.
The impact to the subsequent processing can be canceled by applying $\textbf{P}^{-1}$ before the next layer so the functionality of the DNN remains intact.
This \emph{neuron permutation attack} is examplified in Fig.~\ref{figure:threat}.
One solution to this threat is designing verifier modules that are invariant to the permutation of its inputs, which is challenging due to the loss of information and detached from all established white-box DNN watermarking schemes.
Instead, it is desirable that we can recognize $\textbf{P}$ and cancel its influence by aligning the neurons into their original order.
To perform aligning, we encode neurons by their scalar outputs, which are invariant under any permutation in precedent layers.
The neurons' outputs on training data, which are supposed to be the most diversified, are clustered into several centroids as signals.
To get rid of the deviation from a neuron's normal outputs to the centroids, some trigger samples are generated to correctly evoke these signal outputs as a neuron's identifier code.
To guarantee robust reconstruction, such encoding also needs to have good error-correcting ability against model tuning.

\section{The Proposed Method}
\label{sec:3}
Assume that the watermarked layer contains $N$ homogeneous neurons, the code for a neuron is its outputs on a specialized collection of inputs, known as \emph{triggers}.
Given the triggers, the owner can obtain the codes for neurons and align them properly.
What remains to be specified is the encoding scheme and the generation of triggers.

\subsection{Neuron encoding}
\begin{figure}[htbp]
\centering
\includegraphics[height=5.8cm]{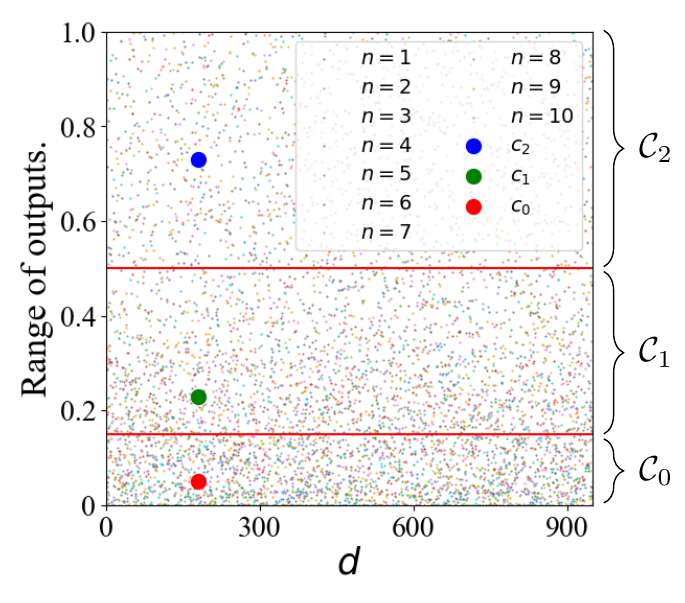}
\caption{Splitting the output space, $D$=1024, $N$=10, and $K$=3.}
\label{figure:encode}
\end{figure}

Denote the length of the code by $T$ and the size of the alphabet by $K$.
Each trigger invokes one output from each neuron and is mapped into one position in each neuron's code, so $T$ is also the number of triggers.
Denote the output of the $n$-th neuron in the watermarked layer for an input $\textbf{x}$ as $y_{n}(\textbf{x})$, let the training dataset be $\left\{\textbf{x}_{d} \right\}_{d=1}^{D}$.
The normal output space of neurons in the watermarked layer is split into $K$ folds.
The centroid of the $k$-th fold, $c_{k}$, is computed as:
\begin{equation}
\label{equation:centroids}
c_{k}=\frac{\sum_{n=1}^{N}\sum_{d=1}^{D}y_{n}(\textbf{x}_{d})\cdot \mathbb{I}[y_{n}(\textbf{x}_{d})\in \mathcal{C}_{k}]}{\lceil ND / K\rceil},
\end{equation}
where $\mathcal{C}_{k}$ is the range of the $k$-th fold containing the $\lceil \frac{NDk}{K}\rceil $-th to the $\lceil \frac{ND(k+1)}{K}\rceil$-th smallest elements in $\left\{y_{n}(\textbf{x}_{d}) \right\}_{d=1,n=1}^{D,N}$.
This process is demonstrated in Fig.~\ref{figure:encode}.

Having determind the centroids, the $n$-th neuron is assigned a code $\textbf{r}_{n}\in \left\{0,1,\cdots,K \right\}^{T}$, the dictionary $\left\{ \textbf{r}_{n}\right\}_{n=1}^{N}$ is spanned using a error correction coding scheme~\cite{9418432}.
It is expected that the output of the $n$-th neuron on the $t$-th trigger, $y_{n}(\textbf{x}_{t})$, is close to $c_{\textbf{r}_{n,t}}$.
To enable error correcting within fewer than $T_{\text{corrupted}}$ positions and each position shifts within at most $K_{\text{corrupted}}$ folds, it is necessary that $T$ and $K$ satistisfy:
\begin{equation}
\label{equation:bound}
N\cdot\left(\sum_{t=1}^{T_{\text{corrupted}}}\binom{T}{t}\cdot K_{\text{corrupted}}^{t} \right)\leq K^{T}.
\end{equation}

\begin{figure*}[!t]
\centering
\subfigure[Generating the $t$-th trigger, the code for the $t$-th position is \texttt{(1402)}.]{
\includegraphics[height=4cm]{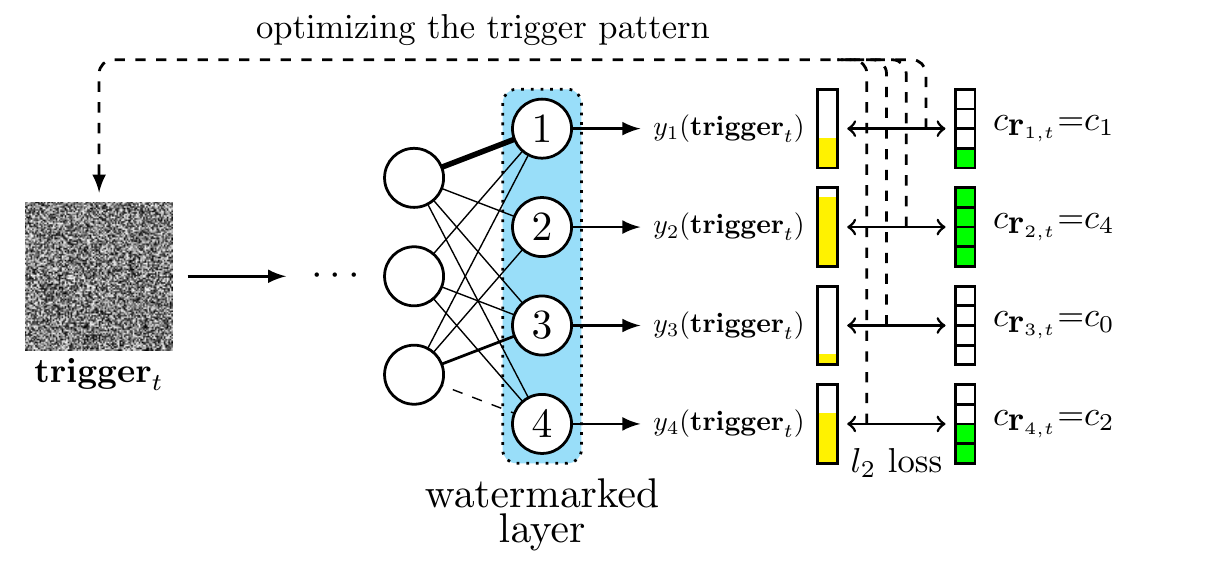}
}\subfigure[Aligning neurons from the intermediate output code \texttt{(1024)} on the $t$-th trigger.]{
\includegraphics[height=4cm]{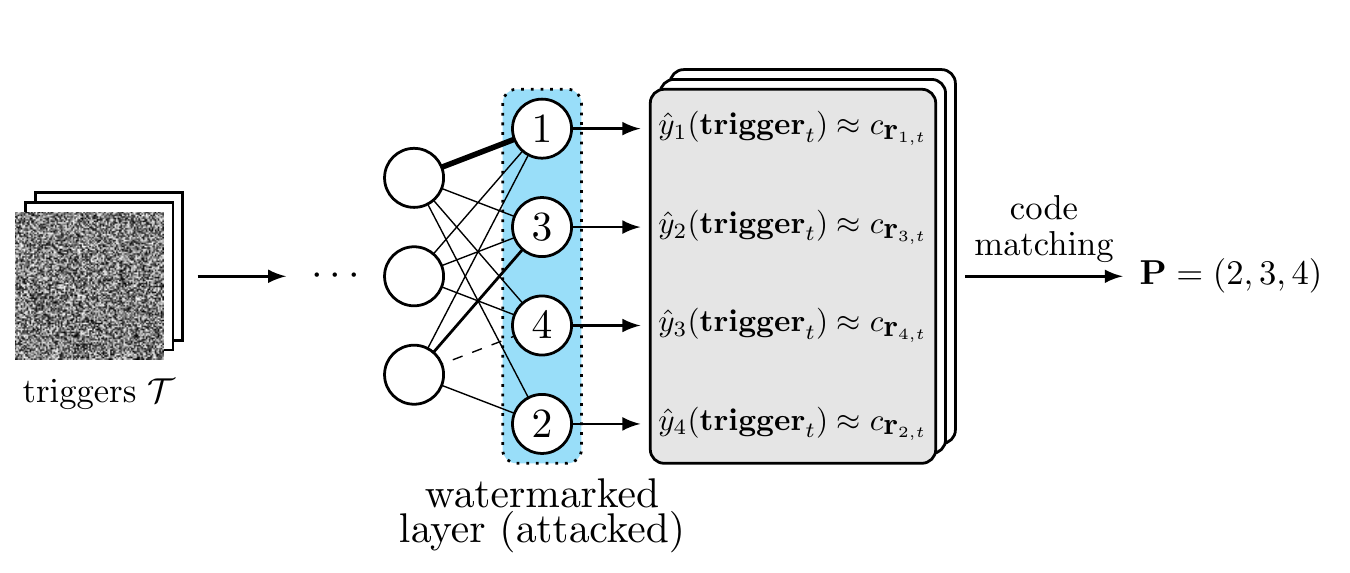}
}
\caption{The trigger generation process and the neuron alignment process.}
\label{figure:triggers}
\end{figure*}

\subsection{Trigger generation}
To generate triggers that correctly evoke the neurons' outputs as codes, we adopt the method in forging adversarial samples~\cite{goodfellow2014explaining}.
Concretely, the $t$-th trigger $\textbf{trigger}_{t}$ is obtained by minimizing the following loss:
\begin{equation}
\label{equation:loss1}
\mathcal{L}_{1}(\textbf{trigger}_{t})=\sum_{n=1}^{N}\|y_{n}(\textbf{trigger}_{t})-c_{\textbf{r}_{n,t}}\|_{2}^{2},
\end{equation}
in which the parameters of the entire DNN are frozen.
To increase the robustness of this encoding against the adversary's tuning, we suggest that $\textbf{t}_{m}$ be optimized w.r.t. the adversarially tuned versions of the watermarked DNN as well.
Let $\left\{y_{n}^{j}\right\}_{j=0}^{J}$ denotes the mapping introduced by the $n$-th neuron under all $J$ kinds of tuning ($j$=0 represents the original model), the loss function becomes:
\begin{equation}
\label{equation:loss2}
\mathcal{L}_{2}(\textbf{trigger}_{t})=\sum_{j=0}^{J}\sum_{n=1}^{N}\|y_{n}^{j}(\textbf{trigger}_{t})-c_{\textbf{r}_{n,t}}\|_{2}^{2},
\end{equation}

The collection of all triggers $\mathcal{T}=\left\{\textbf{trigger}_{t} \right\}_{t=1}^{T}$ forms the owner's evidence for neuron alignment.
This process is demonstrated in Fig.~\ref{figure:triggers} (a).

\subsection{Neuron alignment}
Given the white-box access to the suspicious DNN, the owner can recover the order of neurons in the watermarked layer by the following steps:
(i) Inputting all triggers $\mathcal{T}$ sequentially into the DNN.
(ii) Recording the outputs of the $n$-th neurons in the watermarked layer as $\left\{\hat{y}_{n}(\textbf{trigger}_{t})\right\}_{t=1}^{T}$.
(iii) Transcripting $\hat{y}_{n}(\textbf{trigger}_{t})$ into a code $\hat{\textbf{r}}_{n}\in\left\{0,1,\cdots,K\right\}^{T}$:
\begin{equation}
\nonumber
\hat{\textbf{r}}_{n,t}=\arg\min_{k}\left\{\|\hat{y}_{n}(\textbf{trigger}_{t})-c_{k} \|_{2} \right\}.
\end{equation}
(iv) Transcripting $\hat{\textbf{r}}_{n}$ into an index $i_{n}$:
\begin{equation}
\nonumber
i_{n}=\arg\min_{n'}\left\{\sum_{t=1}^{T}|\textbf{r}_{n',t}-\hat{\textbf{r}}_{n,t}| \right\},
\end{equation}
Finally, the owner aligns all neurons according to their indices and conducts OV using its white-box watermark verifier.
This process is demonstrated in Fig.~\ref{figure:triggers} (b).

\textbf{Remark:} An adaptive adversary might breach this alignment by rescaling the weights across layers.
This can be neutralized by normalizing parameters before alignment or adopting a smaller $K$ to ensure distinguishability.

\section{Experiments and Discussions}
\label{sec:4}
\subsection{Settings}
To examine the validity of the proposed framework, we selected two DNN structures, ResNet-18 and ResNet-50~\cite{he2016deep}.
In each DNN, we selected the second ($l_{2}$) and the third ($l_{3}$) layers to be watermarked.
$l_2$ contains 64 homogeneous neurons and $l_3$ contains 128 ones.
For these convolutional layers, the output where neurons are recognized and decoded is the value of one specific pixel.
Both networks were trained for three computer vision tasks: MNIST~\cite{deng2012mnist}, FashionMNIST~\cite{xiao2017fashion}, and CIFAR-10~\cite{krizhevsky2009learning}.
The training of all DNN backbones and triggers was implemented by Adam~\cite{zhang2018improved} with \texttt{PyTorch}.

\subsection{The configuration of parameters}
\begin{figure}[!t]
\centering
\subfigure[ResNet-18.]{
\includegraphics[width=4.2cm]{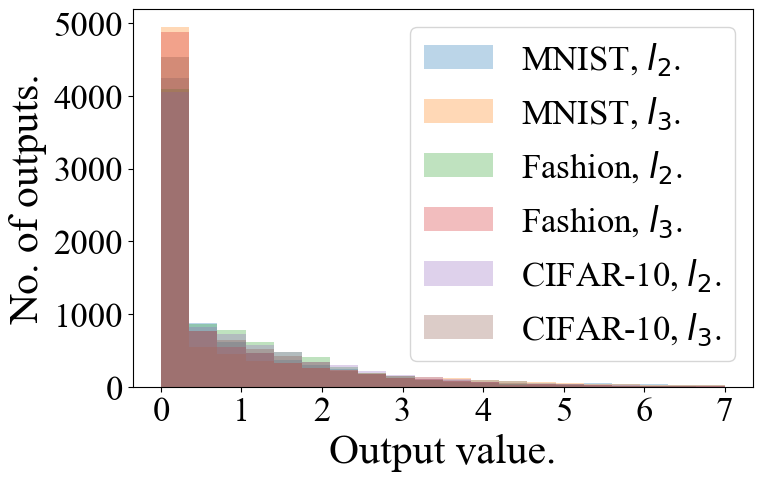}
}\subfigure[ResNet-50.]{
\includegraphics[width=4.2cm]{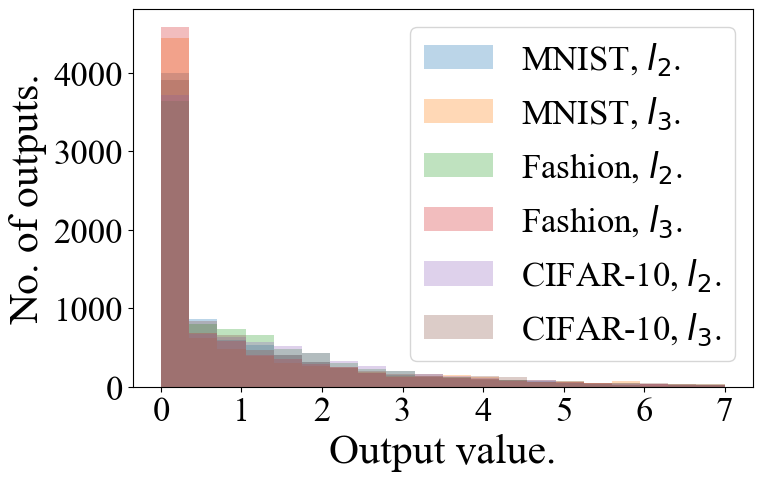}
}
\caption{Distributions of watermarked layers' outputs.}
\label{figure:normal_intermediate}
\end{figure}

To compute the centroids, we measured the distributions of outputs for watermarked layers on normal samples, results are demonstrated in Fig.~\ref{figure:normal_intermediate}.
These distributions remained almost invariant to the selection of dataset, network, and layer.
To ensure maximal distinguishability, we adopted $K$=2 and computed the centroids $c_{0}$=0, $c_{1}$=2.5 by~\eqref{equation:centroids}.
With $K_{\text{corrupted}}$=1, the error correcting ability computed from~\eqref{equation:bound} is shown in Table.~\ref{table:M'}.
We adopted $T$=160 in the following experiments, where flipped bits up to 40\% would not compromise the unique decoding.
\begin{table}[!t]
\centering
\begin{tabular}{c|c|c|c|c|c|c|c|c}
\toprule
\diagbox [width=4em]{$N$}{$T$} & 20 & 40 & 60 & 80 & 100 & 120 & 140 & 160 \\
\midrule[1pt]
64 & 4 & 12 & 21 & 29 & 38 & 47 & 56 & 65 \\
\midrule
128 & 4 & 11 & 20 & 28 & 37 & 46 & 55 & 64 \\
\bottomrule
\end{tabular}
\caption{The maximal number of flipped positions that can be corrected, $T_{\text{corrupted}}$, w.r.t. $N$ and $T$, $K=2$.}
\label{table:M'}
\end{table}

\subsection{Comparative studies}
For comparison, we compared five candidate schemes for trigger selection. \textbf{(N)}: \textbf{N}ormal samples from the training dataset.
\textbf{(R)}: \textbf{R}andom noises.
\textbf{(O)}: \textbf{O}ut-of-dataset samples.
\textbf{(T1)}: \textbf{T}riggers generated be minimizing $\mathcal{L}_{\textbf{1}}$ in~\eqref{equation:loss1}.
\textbf{(T2)}: \textbf{T}riggers generated by minimizing $\mathcal{L}_{\textbf{2}}$ in~\eqref{equation:loss2} with $J$=6 involving three rounds of fine-tuning and three of neuron-pruning.
For \textbf{(N)(R)(O)}, the centroids were also selected by~\eqref{equation:centroids} and the code of each neuron at the $t$-th position was assigned as the index of the closest centroid to its output on the $t$-th input.

\begin{figure}[!t]
\centering
\subfigure[$l_2$.]{
\includegraphics[width=4.3cm]{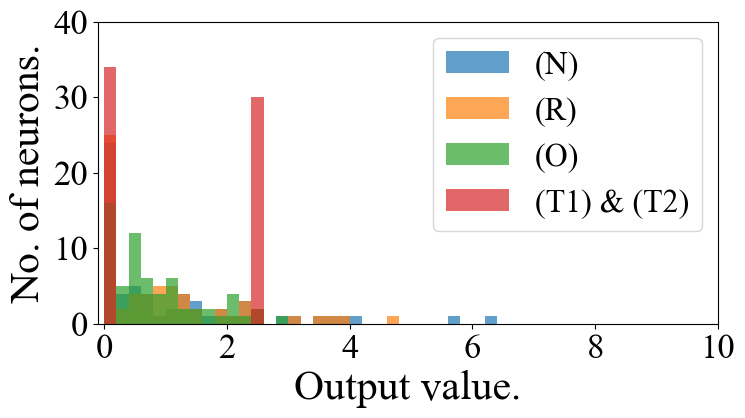}
}\subfigure[$l_3$.]{
\includegraphics[width=4.3cm]{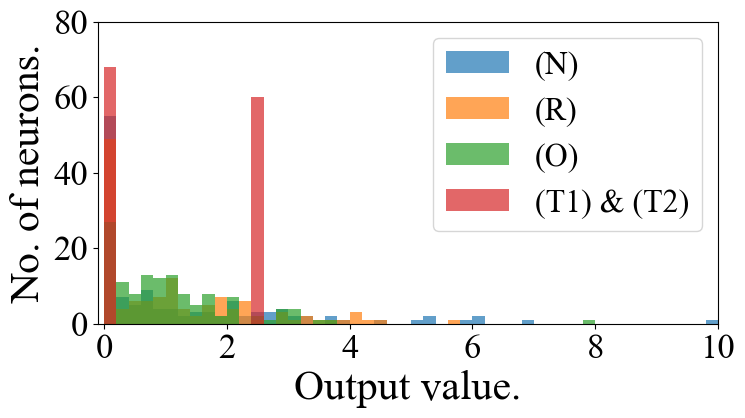}
}
\centering
\subfigure[$l_2$.]{
\includegraphics[width=4.3cm]{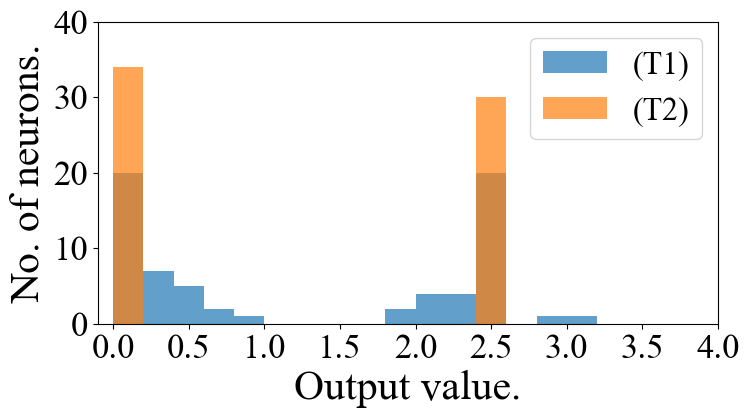}
}\subfigure[$l_3$.]{
\includegraphics[width=4.3cm]{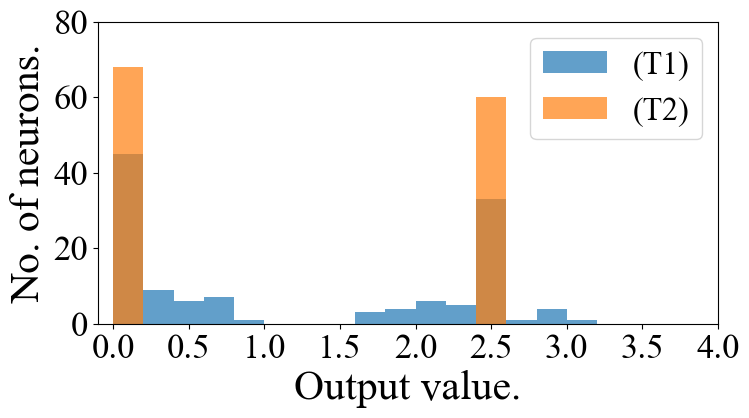}
}
\caption{The distribution of watermarked layers' outputs for different triggers. In (c)(d), the DNN has been fine-tuned.}
\label{figure:advantage}
\end{figure}

The outputs of neurons in ResNet-50 trained on CIFAR-10 for one input are shown in Fig.~\ref{figure:advantage}(a)(b).
From which we noticed that the outputs w.r.t. \textbf{(T1)(T2)} concentrated to $K$=2 centers.
(The percentage of neurons outputting approximately 0 or 2.5 was not strictly 50\%, since the outputs of around 2\% of neurons were uniformly zero.)
Therefore, the code of neurons under \textbf{(T1)(T2)} can be unambiguously retrieved.
\begin{table}[!t]
\centering
\begin{tabular}{c|c|c|c|c|c}
\toprule
\textbf{Metric} \ & \textbf{(N)} & \textbf{(R)} & \textbf{(O)} & \textbf{(T1)} & \textbf{(T2)} \\
\midrule[1pt]
Inter-cluster ($\uparrow$). & 2.4 & 1.9 & 1.5 & \textbf{2.5} & \textbf{2.5} \\
\midrule
Intra-cluster ($\downarrow$). & 1.3 & 0.8 & 0.8 & \textbf{0.1} & 0.2 \\
\midrule
Accuracy (\%). & 1.0 & 2.3 & 1.3 & \textbf{98.4} & 97.2 \\
\bottomrule
\end{tabular}
\caption{The statistics of neurons' outputs.}
\label{table:inandout}
\end{table}
Numerically, we computed the averaged inter/intra-cluster distance for all trigger patterns with two clusters obtained by $K$-means~\cite{li2019bayesian} and the accuracy of aligning against random shuffling on $l_3$, the results are listed in Table.~\ref{table:inandout}.
From which we justified that the codes derived by \textbf{(T1)(T2)} are more informative.
After fine-tuning, the distributions of outputs under \textbf{(T1)(T2)} were differentiated as shown in Fig.~\ref{figure:advantage}(c)(d), so \textbf{(T2)} is more robust against model tuning.

\subsection{The performance of watermarking backends}
To study the performance of white-box DNN watermarking schemes after the neuron permutation attack and alignment, we considered four state-of-the-art watermarking schemes: Uchida~\cite{uchida2017embedding}, Fan~\cite{fan2021deepip}, Residual~\cite{liu2021watermarking}, and MTLSign~\cite{ours}.
All watermarks were embedded into both $l_{2}$ and $l_{3}$.

We conducted three attacks to the watermarked layer:
\textbf{(NP)}: \textbf{N}euron \textbf{P}ermutation;
\textbf{(FTP)}: \textbf{F}ine-\textbf{T}uning and neuron \textbf{P}ermutation;
\textbf{(NPP)} \textbf{N}euron-\textbf{P}runing and \textbf{P}ermutation.
Then we applied neuron alignment and recorded the percentage of correct verifications from the watermarking backends in 1,000 instances, results are summarized in Table.~\ref{table:backends}.
Without neuron alignment, any permutation-based attack can reduce the OV accuracy to 0.0\%.
After alignment, the accuracy in all cases increased significantly.
Compared with \textbf{(T1)}, \textbf{(T2)} is more robust against tuning and pruning in better reconstructing the order of neurons.
Therefore, by adopting the neuron alignment framework, the security levels of these watermarking schemes are substantially increased.
Meanwhile, the trigger generation process does not modify the original DNN, so it can be parallelized and would not bring extra damage to the protected DNN.

\begin{table}[!t]
\centering
\begin{tabular}{m{0.85cm}<\centering|m{1.4cm}<\centering|m{1.4cm}<\centering|m{1.4cm}<\centering|m{1.4cm}<\centering}
\toprule
\textbf{Attack} & Uchida & Fan & Residual & MTLSign \\
\midrule[1pt]
\textbf{(NP)} & \tabincell{c}{0.0,\\ (95.7,95.5)} & \tabincell{c}{0.0,\\ (95.1,95.1)} & \tabincell{c}{0.0,\\ (98.3,98.0)} & \tabincell{c}{0.0,\\ (99.1,98.6)} \\
\midrule
\textbf{(FTP)} & \tabincell{c}{0.0,\\ (69.4,90.3)} & \tabincell{c}{0.0,\\ (74.3,75.2)} & \tabincell{c}{0.0,\\ (82.7,87.4)} & \tabincell{c}{0.0,\\ (79.9,87.9)} \\
\midrule
\textbf{(NPP)} & \tabincell{c}{0.0,\\ (54.4,74.3)} & \tabincell{c}{0.0,\\ (67.9,76.5)} & \tabincell{c}{0.0,\\ (77.7,82.6)} & \tabincell{c}{0.0,\\ (75.4,83.9)} \\
\bottomrule
\end{tabular}
\centering
\begin{tabular}{m{0.85cm}<\centering|m{1.4cm}<\centering|m{1.4cm}<\centering|m{1.4cm}<\centering|m{1.4cm}<\centering}
\toprule
\textbf{Attack} & Uchida & Fan & Residual & MTLSign \\
\midrule[1pt]
\textbf{(NP)} & \tabincell{c}{0.0,\\ (96.3,95.4)} & \tabincell{c}{0.0,\\ (96.0,95.6)} & \tabincell{c}{0.0,\\ (99.1,99.4)} & \tabincell{c}{0.0,\\ (98.7,98.7)} \\
\midrule
\textbf{(FTP)} & \tabincell{c}{0.0,\\ (70.1,89.1)} & \tabincell{c}{0.0,\\ (74.9,75.9)} & \tabincell{c}{0.0,\\ (84.6,88.0)} & \tabincell{c}{0.0,\\ (79.6,90.5)} \\
\midrule
\textbf{(NPP)} & \tabincell{c}{0.0,\\ (58.9,72.5)} & \tabincell{c}{0.0,\\ (63.7,73.4)} & \tabincell{c}{0.0,\\ (79.8,81.3)} & \tabincell{c}{0.0,\\ (75.9,84.9)} \\
\bottomrule
\end{tabular}
\caption{The performance of watermarking backends after neuron alignments for ResNet-18 and ResNet-50.
The results in each entry are: the accuracy of OV after the attack and its increase (in \%) with alignment by (\textbf{(T1)}, \textbf{(T2)}), averaged across three datasets. }
\label{table:backends}
\end{table}

\section{Conclusions}
\label{sec:5}
We propose a neuron alignment framework to enhance established white-box DNN watermarking schemes.
Clustering and error-correcting encoding are adopted to ensure the availability and distinguishability of neuron encoding.
Then we use a generative method to forge triggers that can correctly and robustly reveal the neurons' order.
Experiments demonstrate the effectiveness of our framework against the neuron permutation attack, a realistic threat to OV for DNN.

\bibliographystyle{IEEEbib}
\bibliography{WM.bib}

\end{document}